\documentclass[twocolumn,showpacs,preprintnumbers,prb,amsmath,amssymb]{revtex4}

\usepackage{graphicx}
\usepackage{dcolumn}
\usepackage{bm}
\usepackage{color}

\begin{document}


\title{Magnetic and transport properties of rare-earth-based half-Heusler phases \textit{R}PdBi: \\prospective systems for topological quantum phenomena}

\author{K. Gofryk$^{1,2}$}
\email{gofryk@lanl.gov}
\author{D.~Kaczorowski$^{2}$}
\author{T. Plackowski$^{2}$}
\author{A.~Leithe-Jasper$^{3}$}
\author{Yu.~Grin$^{3}$}

\affiliation{$^{1}$Condensed Matter and Thermal Physics, Los Alamos National Laboratory, Los Alamos, New Mexico 87545, USA\\$^{2}$Institute of Low Temperature and
Structure Research, Polish Academy of Sciences, P.O. Box 1410,
50-950 Wroc{\l}aw, Poland\\$^{3}$Max-Planck-Institut f{\"{u}}r
Chemische Physik fester Stoffe, N{\"{o}}thnitzer Str. 40, 01187 Dresden, Germany}


\begin{abstract}
\textit{R}PdBi (\textit{R} = Er, Ho, Gd, Dy, Y, Nd) compounds were studied by means of x-ray diffraction, magnetic
susceptibility, electrical resistivity, magnetoresistivity, thermoelectric power and Hall effect measurements, performed in the temperature range 1.5-300~K and in magnetic fields up to
12~T. These ternaries, except diamagnetic
YPdBi, exhibit localized magnetism of $R^{3+}$ ions, and order antiferromagnetically at low temperatures ($T_{N}$~=~2-13~K). The transport measurements revealed behavior characteristic of semimetals or narrow-band semiconductors. Both, electrons and holes contribute to the conductivity with dominant role of $p$-type carriers. The Hall effect of ErPdBi is strongly temperature and magnetic field dependent, reflecting complex character of the underlying electronic structures with multiple electron and hole bands. \textit{R}PdBi, and especially DyPdBi, exhibit very good thermoelectric properties with a power factor coefficient $PF$ ranging from 6 to 20~$\mu$Wcm$^{-1}$K$^{-2}$.

\end{abstract}

\pacs{}
\maketitle

\section{Introduction}

The rare-earth (R)-based ternary compounds with
the composition \textit{RTX}, where $T$ denotes a $d$-electron transition metal and $X$ is a $p$-element, have been studied extensively during last years\cite{1}. Specifically, phases crystalizing in a cubic MgAgAs-type of crystal structure which are also known under the very unlucky name "half-Heusler" phases, are well known because of many different intriguing and extraordinary physical properties observed in this family\cite{f1}, including half-metallic\cite{2} and semiconducting-like behavior\cite{3,f2,KGprb2}, giant magnetoresistivity\cite{4,5} or heavy fermion state.\cite{6,7,8} For this reason these phases have been named "compounds with properties on request"\cite{pierre}. Very recently, based on first-principle calculations, a topological insulating state has been predicted in heavy metal-element containing half-Heusler phases\cite{NM1,NM2}. The topological insulator is a new state of quantum matter having a full insulating gap in the bulk, but with topologically protected gapless surface or edge states on the boundary. It has been demonstrated that some compounds from the series \textit{R}AuPb, \textit{R}PtSb, \textit{R}PtBi or \textit{R}PdBi exhibit a band inversion similar to that discovered in HgTe\cite{NM1,NM2}. Band structure calculations show that these phases should possess a semimetallic or narrow-band semiconducting state and the topological state may be created by applying strain (such as crystallographic distortion) or by designing an appropriate quantum-well structure, similar to the case of HgTe. Interestingly, since many of those \textit{MgAgAs} compounds contain a rare-earth element, additional properties ranging from superconductivity to magnetism and heavy fermion behavior may be realized within a topological matrix. These properties can open new research directions in realizing the quantized anomalous Hall effect\cite{n5,n6}, topological superconductors with Majorana fermion excitations\cite{n7} and/or image magnetic monopole effect.\cite{n9}
In this paper we present the magnetic and electrical transport properties of a series of rare-earth-based \textit{R}PdBi phases. The compounds are either paramagnetic or antiferromagnetic with Ne\'{e}l temperatures of several kelvin (except of YPdBi which shows diamagnetic behavior) due to well localized 4\textit{f} electrons. The transport measurements revealed semimetallic behavior with likely dominance of $p$-type carriers. The electronic structure has complex character with the presence of multiple electron and hole bands. Moreover, this family of compounds seems to possess very good thermoelectrical parameters, prospective for practical applications.

\section{Experimental details}

Polycrystalline samples of ErPdBi, HoPdBi, GdPdBi, YPdBi, DyPdBi and NpPdBi were prepared by arc-melting the constituents (\textsl{RE}: 99.9 \%wt, Pd: 99.999 \%wt, Sb: 99.999 \%wt) under ultra pure argon atmosphere. Since additional heat treatment led to multiphase products, the as-cast
samples were used for physical measurements. The quality of the obtained materials was
checked at 300~K by powder x-ray diffraction (Huber Guinier G670
image plate camera with CuK$\alpha_{1}$ radiation, \textit{$\lambda$} =
1.5406 \AA, and silicon as an internal standard, $a$ = 5.43119 \AA), optical metallography (Zeiss Axioplan 2 optical microscope with a CCD
camera) and EDX analysis (Philips XL30 scanning electron microscope with
integrated EDXS system and S-UTW-Si-(Li) detector). All these studies have shown that the samples studied were single phase with a cubic structure of the MgAgAs-type (space group $F\overline{4}$3$m$). This crystal structure is shown in Fig.~1, together with a unit cell adopted by so-called Heusler phases (MnCu$_{2}$Al-type; space group $Fm\overline{3}m$). The lattice parameters obtained from the x-ray diffraction data are: $a$~=~6.5952, 6.6012, 6.6343, 6.6391, 6.6906 and 6.7257~{\AA} for ErPdBi, HoPdBi, DyPdBi, YPdBi, GdPdBi and NdPdBi,  respectively. The derived values are in a good agreement with those given in the literature\cite{21d}.

\begin{figure}[t]
\centering
\includegraphics[width=0.4\textwidth]{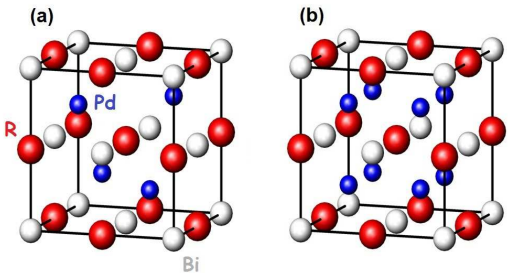}
\caption{(color online) Crystal structure of (a) MgAgAs- and (b) MnCu$_{2}$Al-type phases.}\label{str}
\end{figure}

Magnetization measurements were carried out in the temperature range 1.7-300~K and in magnetic fields up to 5~T using a Quantum Design MPMS-5 magnetometer. The
electrical resistivity was measured from 4 to 300~K by a standard
four-point $DC$ technique. The Hall effect
measurements were performed in the temperature range
1.5-300 K and in applied fields up to 12 T (Oxford Instruments
TESLATRON). The thermoelectric power was studied
from 6 to 300~K, employing a home-built setup using pure copper as a
reference material.

\section{Results and discussion}

The magnetic measurements revealed that HoPdBi, DyPdBi, GdPdBi and NdPdBi order antiferromagnetically at 2, 3.5, 13 and 4 K, respectively, while ErPdBi remains paramagnetic down to the lowest temperature studied (see Fig.~2a). In wide temperature ranges the $\chi^{-1}(T)$ exhibit linear temperature dependence. They may be well described by the Curie-Weiss law and the effective magnetic moments $\mu_{B}$ being close to those expected for free $R^{3+}$ ions within
Russell-Saunders coupling scenario, $\mu_{eff}^{R^{3+}} = g[J(J+1)]^{1/2}$. This indicates good localization of the magnetic moments on the rare-earth atoms. Moreover, the obtained values of the paramagnetic Curie temperature are small and negative, consistent with the type of magnetic ordering in these systems. As the only exception, YPdBi, which does not contain $4f$ electrons, shows a diamagnetic behavior (see Fig.~2b) that may reflect the formation of a time-reversal-invariant spin-orbit ground state. It is worth to recall that it is one of the conditions for realizing a $Z_{2}$ = -1 topological insulating state.\cite{NM2}\\

\begin{figure}[t]
\centering
\includegraphics[width=0.45\textwidth]{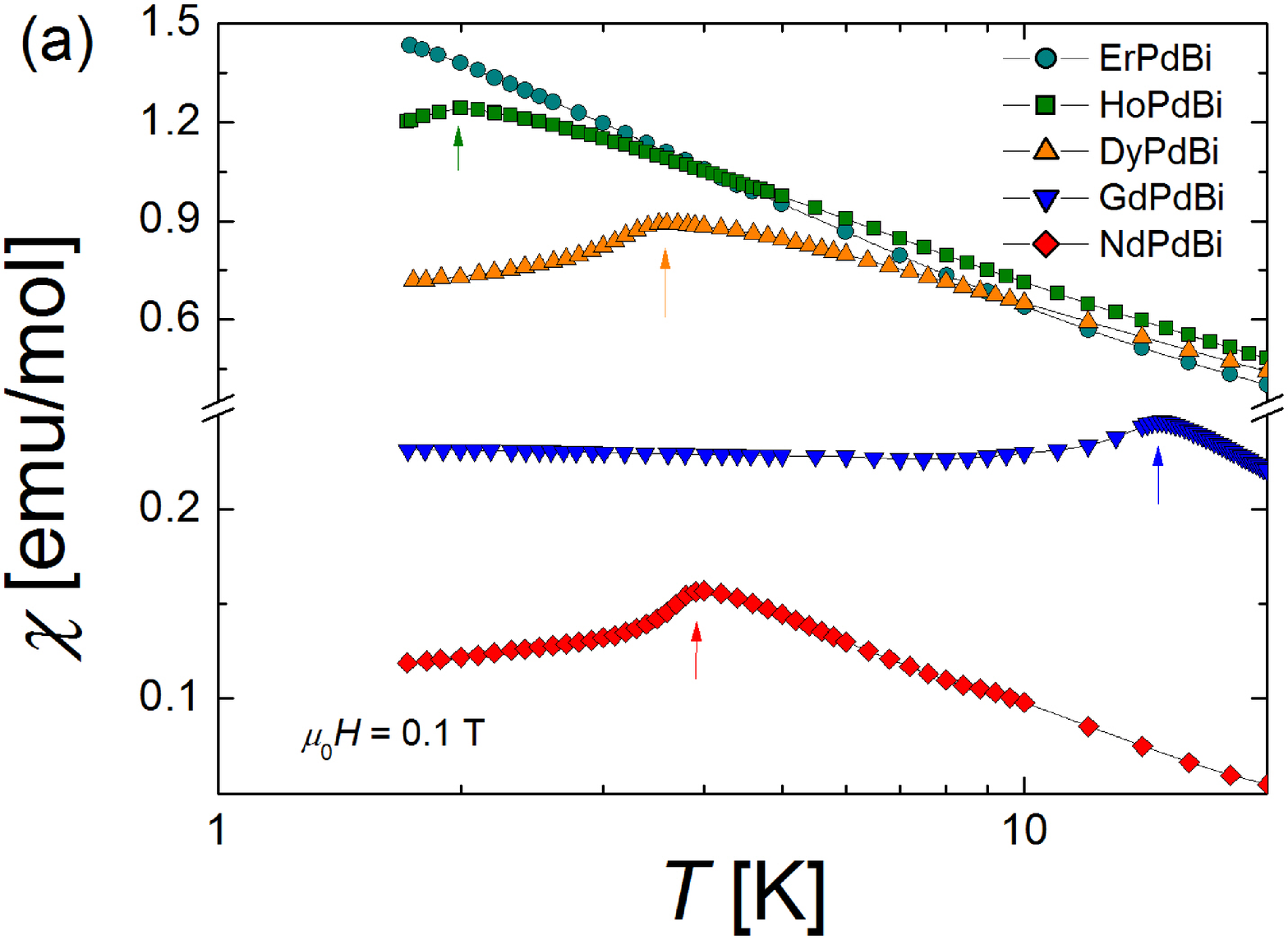}
\includegraphics[width=0.45\textwidth]{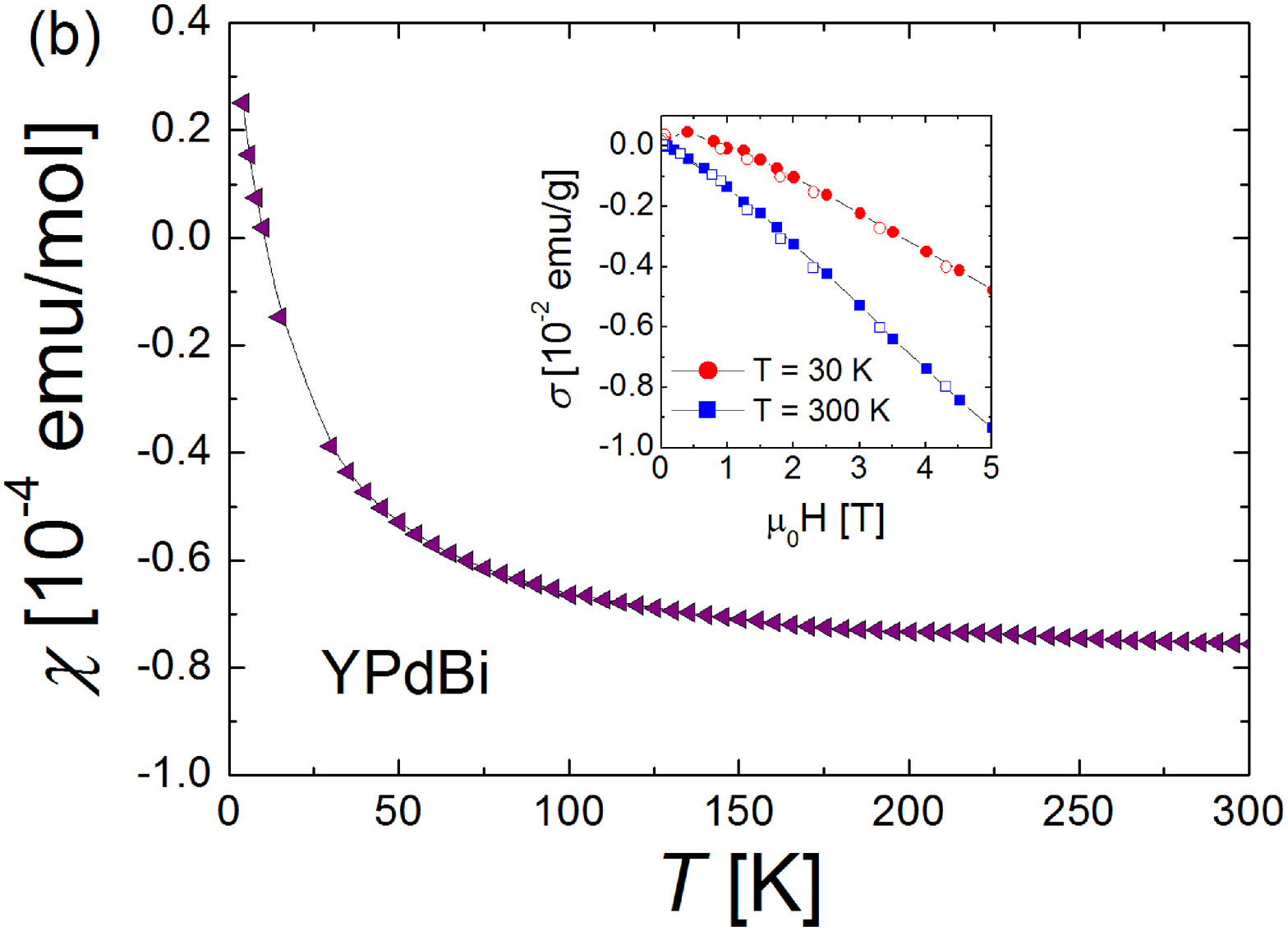}
\caption{(color online) (a) Low temperature dependencies of the magnetic susceptibility of \textit{R}PdBi compounds. Arrows mark the antiferromagnetic phase transition. (d) Temperature dependence of the magnetic susceptibility of YPdBi. Inset: magnetic field variation of the magnetization in YPdBi taken at 30 and 300~K with increasing (solid symbols) and decreasing (empty symbols) field.}\label{sus}
\end{figure}

The temperature dependencies of the electrical resistivity of the \textit{R}PdBi phases are shown in Fig.~3. In general, the magnitude and the temperature variations of the resistivity are characteristic of semimetals or narrow-gap semiconductors\cite{dornhaus}. In the case of ErPdBi, GdPdBi, NdPdBi and DyPdBi, the $\rho(T)$ exhibits two different regimes. At high temperatures the electrical resistivity displays a semiconducting-like character ($d\rho/dT~<~0$), signaling carrier excitations over a small energy gap $E_{g}$ near the Fermi level. This behavior is followed at lower temperatures by a metallic-like dependence of $\rho(T)$ ($d\rho/dT~>~0$). In general, the observed behavior is reminiscent of those typical for doped semiconductors where, due to atomic disorder, defects and/or wrong stoichiometry some donor or acceptor levels are present, leading to metallic-like $\rho(T)$. In order to account for this rather complex temperature behavior a simple model of the electronic band structure in narrow gap semiconductors may be used (see Refs.\onlinecite{st,KGprb1,Jmmm,KGprb2}). Recently, this model has been successfully applied to describe the temperature dependence of the resistivity in some filled skutterudites\cite{st} as well as in compounds ErPdSb \cite{KGprb2}, HoPdSb \cite{KGprb1} and DyPdBi \cite{KGprb1}. In this approach one considers two rectangular bands of height $N$, separated by an energy gap $E_{g}$, and the Fermi level situated just below the gap. In order to describe the metallic conductivity at low temperatures, the presence of some temperature-independent amount of carriers $n_{0}$ is assumed. Thus the total number of carriers can be expressed as

\begin{equation}
n(T)=n_{0}+\sqrt{n_{e}(T)n_{h}(T)}
\end{equation}

where the concentrations of electrons $n_{e}(T)$ and holes $n_{h}(T)$
are given by

\begin{equation}
n_{e}(T)=-NE_{g}+Nk_{B}Tln2\left[1+\left(\frac{E_{g}}{k_{B}T}\right)\right]
\end{equation}

\begin{equation}
n_{h}(T)=-Nk_{B}Tln2
\end{equation}

Based on these assumptions and approximations, $\rho(T)$ can be defined as as

\begin{equation}
\rho(T)=\frac{n_{0}\rho_{0}+\rho_{ph}(T)}{n(T)}\label{b}
\end{equation}

In this formula $\rho_{ph}(T)$ represents the phonon contribution to the resistivity, which is further assumed to be  $\rho_{ph}(T) = AT$ for temperatures above $\Theta_{D}$/10, where $\Theta_{D}$ is the Debye
temperature (see also Ref.\onlinecite{KGprb1}).

As is seen in Fig.~3, above $\sim$25~K, the above model provides a quite satisfactory  description of the electrical resistivity of all of the compounds studied (see solid lines in Fig.~3). The obtained parameters are collected in Table II. As may be seen, the number of carriers $n_{0}$ as well as the parameters characterizing the bands and the gap ($N$ and $E_{g}$) nicely reflect the differences between the compounds studied as regards the character and magnitude of their electrical conductivity.

\begin{figure}[t]
\centering
\includegraphics[width=0.45\textwidth]{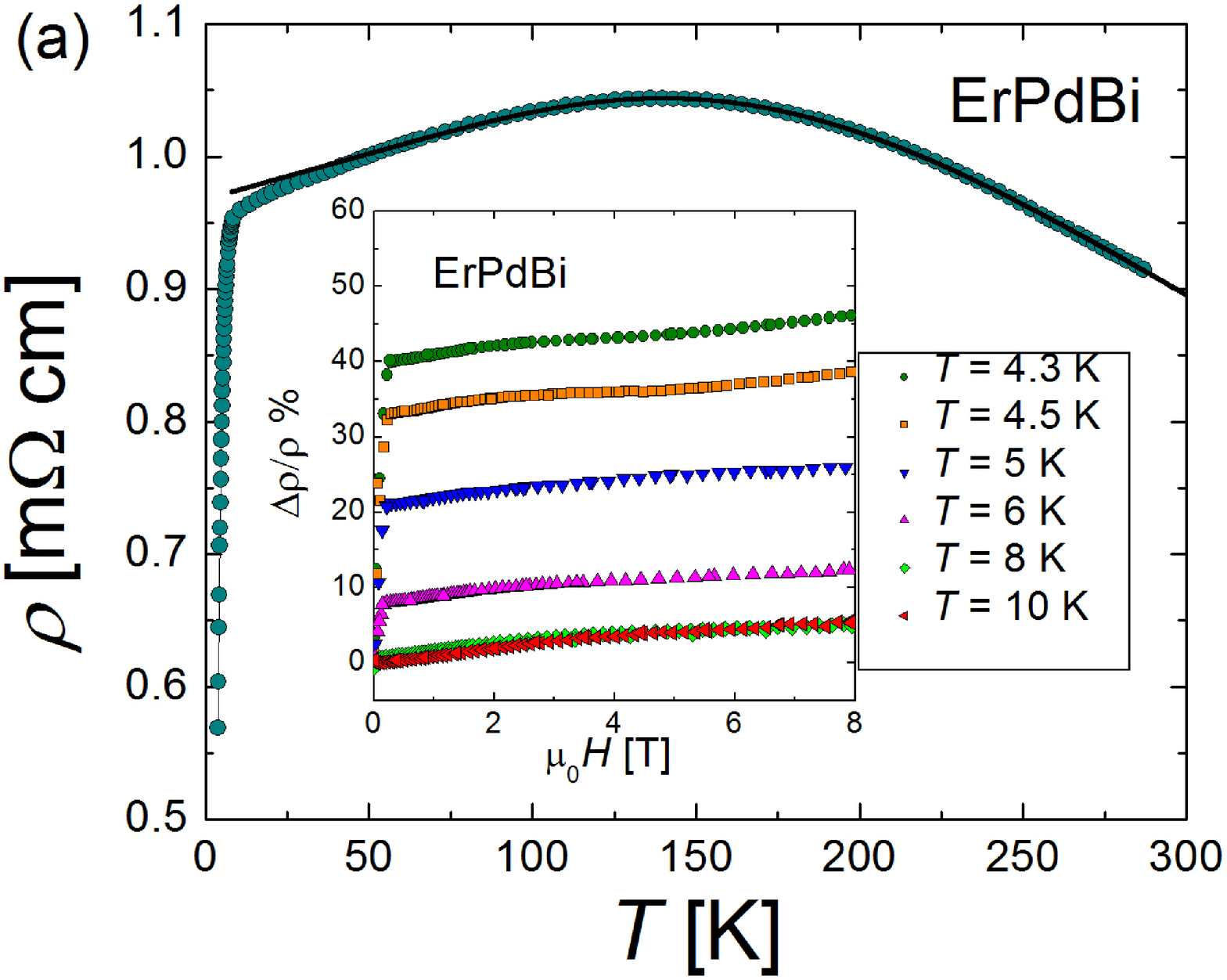}
\includegraphics[width=0.45\textwidth]{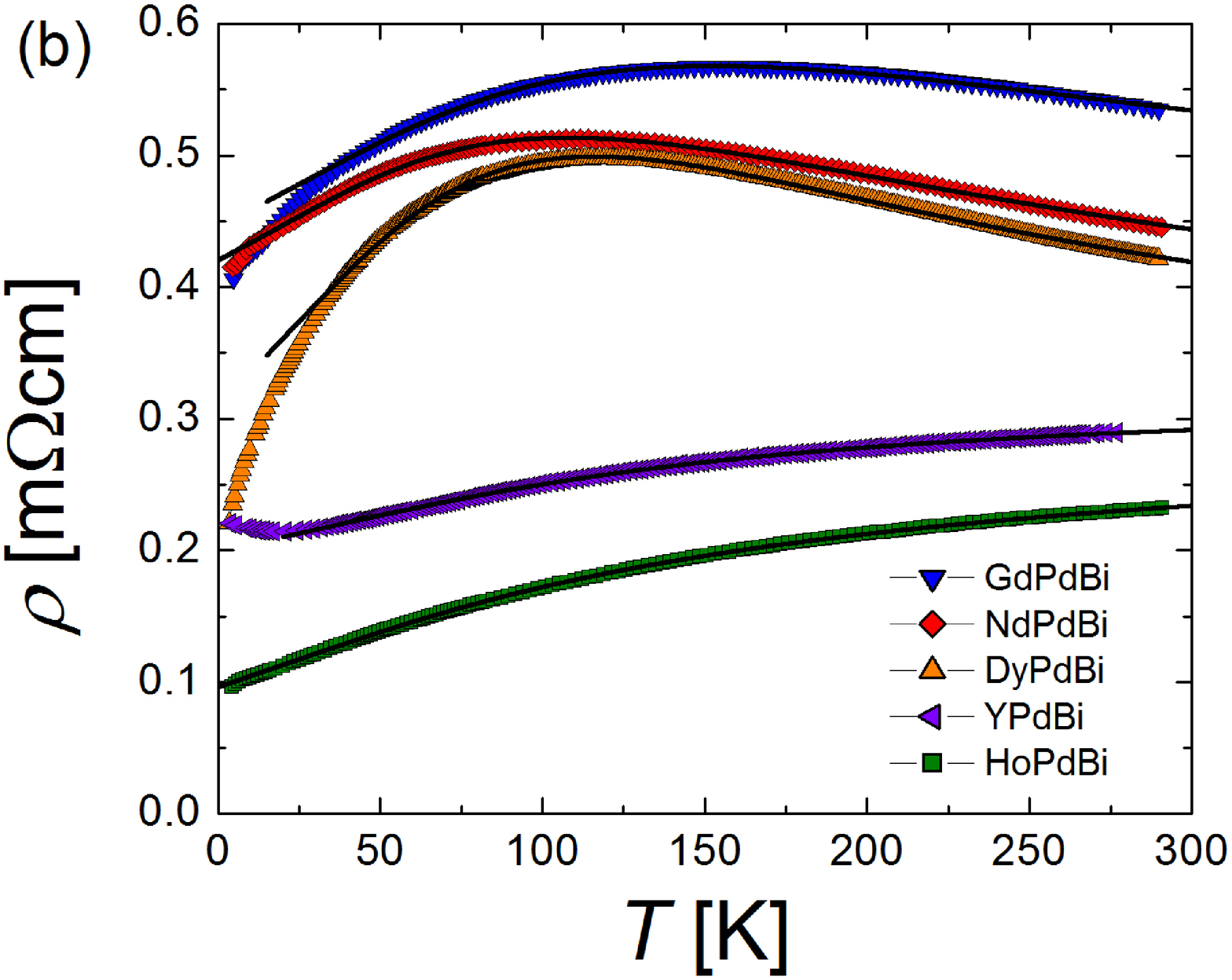}
\caption{(color online) Temperature dependencies of the electrical resistivity of \textit{R}PdBi compounds. The solid lines are the fits of eq.\ref{b} to the experimental data (see text). Inset: magnetoresistivity isotherms vs. field taken for ErPdBi.}\label{zt}
\end{figure}

For ErPdBi a pronounced drop in $\rho(T)$ at about 7~K is observed. This behavior is very similar to that previously observed in the \textit{R}PdSb system, especially for ErPdSb (see Refs.\onlinecite{Jmmm,KGprb1,KGprb2}). As can be inferred from the inset to Fig.~3a, the anomaly in $\rho(T)$ can be suppressed by magnetic field, in a manner typical for superconductivity. However, neither ErPdBi nor ErPdSb are bulk superconductors, as no corresponding features are observed in their magnetic and heat capacity data (see Ref.~\onlinecite{Jmmm} for a more detailed discussion of this issue). Thus, the origin of the unusual low-temperature behavior of the resistivity of ErPdBi is unknown. Similarly, at the present stage we cannot reliably interpret the anomalous feature seen in $\rho(T)$ of YPdBi, i.e. an increase of the resistivity below 19~K.\\

\begin{figure}[t]
\centering
\includegraphics[width=0.45\textwidth]{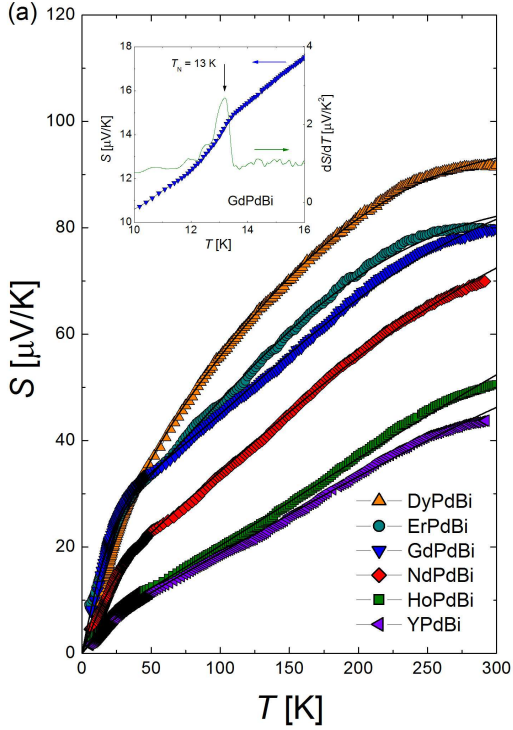}
\caption{(color online) Temperature dependencies of the thermoelectric power of \textit{R}PdBi compounds. The solid lines represent least-square fits of the two band model (eq.\ref{ban}) to the experimental data (see text). Inset shows S(T) and dS/dT of GdPdBi in the vicinity of the magnetic phase transition.}\label{T}
\end{figure}

As the thermopower is a sensitive probe of energy relative to the Fermi level, it can be used as a tool to characterize the electronic structure, especially in the vicinity of the narrow gap or pseudogap. Fig.~4 shows the temperature dependencies of the Seebeck coefficient of the \textit{R}PdBi series. The magnetic phase transitions in these compounds manifest themselves as tiny anomalies in $S(T)$, better seen on the temperature derivative $dS/dT$ (as an example see in the inset in Fig.~4 the behavior of GdPdBi). In general, the overall magnitude and the temperature variations of the thermopower of \textit{R}PdBi are characteristic of low carrier density semimetals\cite{dur,aliev,aliev2}. At 300~K, the thermoelectric power is as large as about 40-90~$\mu$V/K, which corresponds within a single band model to the effective carrier concentration of the order of 10$^{20}$~cm$^{-3}$. For all the compounds studied the Seebeck coefficient is positive in the entire temperature range, thus suggesting that holes may dominate the electrical and heat transport. However, strongly curvilinear character of $S(T)$ hints at the presence of complex electronic structure in the vicinity of the Fermi energy. Indeed, the most recent theoretical calculations have revealed a fairly complex electronic structure in YPdBi (Ref. \onlinecite{NM1}) and in other $RTX$ systems,\cite{NM2} with electron and hole bands close to the Fermi energy.

To account for this complexity, the temperature dependencies of the thermoelectric power of the \textit{R}PdBi compounds were analyzed in terms of a phenomenological model, which describes scattering of carriers on two quasiparticle bands laying close to the Fermi level.\cite{bando} In this approach, it is assumed that the conduction electrons are scattered independently on a narrow $N$ and a wide $W$ bands approximated by Lorentzians. The thermoelectric power in such a system can be expressed by a modified Mott's formula\cite{bando}

\begin{equation}
S(T)=S_{N}(T)+S_{W}(T)=\frac{a_{N}T}{b_{N}^{2}+T^{2}}+\frac{a_{W}T}{b_{W}^{2}+T^{2}}\label{ban}
\end{equation}

where
\begin{equation}
a_{N,W} = \frac{2\Delta_{N,W}}{|e|},
\end{equation}

and
\begin{equation}
b_{N,W}^{2}=\frac{3\Delta_{N,W}^{2}+\Gamma_{N,W}^{2}}{\pi^{2}k_{B}^{2}}.
\end{equation}

In the above equations, $\Delta_{N,W}$ represents the positions of the Lorenzians with relation to the Fermi energy, while $\Gamma_{N,W}$ stands for their width. As shown by the solid lines in Fig.~4 the above model provides a good description of the experimental data of \textit{R}PdBi. The so-obtained parameters, $\Delta_{N,W}$ and $\Gamma_{N,W}$, are listed in Table II.

\begin{table*}[t]
 \centering
 \caption{Transport parameters for $R$PdBi ($R$~=~Er, Ho, Dy, Gd, Nd and Y). $n_{0}$~-~number of carriers at T~=~0~K,
$\rho_{0}$~-~residual resistivity, $N$~-~density of states,
$\rho_{300K}$~-~resistivity measured at T~=~300~K,
$S_{300K}$~-~thermopower measured at T~=~300~K,
$E_{g}$~-~energy gap, $\Delta_{N,W}$ and $\Gamma_{N,W}$~-~position of the narrow and wide bands with relation to the Fermi energy and their width (see text).}
 \vspace{0.5em}
  \begin{tabular}{l c c c c c c c c c c}
  \hline\hline
   Compound & $n_{0}$ & $N$ & $\rho_{300K}$ & $E_{g}$ & $S_{300K}$&$\Delta_{N}$&$\Gamma_{N}$&$\Delta_{W}$&$\Gamma_{W}$ \\
  &[$f.u^{-1}$] &  [eV$^{-1}$] & [$\mu\Omega$cm] & [meV] & [$\mu$V/K] &[meV]&[meV] &[meV]&[meV]\\
  \hline
  ErPdBi& 0.07 &  5.96 & 900 &  94 & 80 & 0.45 & 5.1 & 32 & 52 \\
  HoPdBi& 0.17 &  8.01 & 234 &  32 & 50 & 0.12 & 5.5 & 110 & 130 \\
  DyPdBi& 0.16 &  24.5 & 422 &  35 & 92 & 0.65 & 8.1 & 34 & 47 \\
  YPdBi& 0.27 &  7.85 & 293 &  44.5 & 44 & 0.15 & 7.1 & 72 & 126 \\
  GdPdBi& 0.13 &  8.1 & 533 &  43 & 79 & 0.65 & 5.8 & 44 &65 \\
  NdPdBi& 0.12 & 9.65 & 442 &  34.5 & 69 & 0.35 & 6.6 & 55 &79 \\
  \hline\hline
  \end{tabular}
  \label{2dys111Bi}
\end{table*}

The relatively large values of the Seebeck coefficient at room temperature together with rather small electrical resistivity found for the \textit{R}PdBi materials yield quite enhanced magnitude of the thermoelectric power factor $PF~=~S^{2}/\rho$, ranging from 6.3~$\mu$Wcm$^{-1}$K$^{-2}$ for YPdBi to 20.3~$\mu$Wcm$^{-1}$K$^{-2}$ for DyPdBi. These values are much larger than the ones obtained for \textit{R}PdSb compounds\cite{KGprb2} and comparable with $PF\sim$20-25~$\mu$Wcm$^{-1}$K$^{-2}$ derived for the (Zr,Hf)NiSn system doped by Ta or Nd\cite{hohl}. It is worth emphasizing that a large value of the power factor is the
main prerequisite for good thermoelectrical efficiency, yet thermal
conductivity studies are indispensable to fully characterize the
thermoelectric potential of a given material. The performance of
thermoelectric devices is quantified by a dimensionless figure of
merit $ZT=S^{2}/\rho\kappa$, where $S$ is the Seebeck coefficient,
$\rho$ is the electrical resistivity, and $\kappa$ stands for the
thermal conductivity. Such measurements would be needed especially for
DyPdBi which seems to be the most promising in terms of
thermoelectric performance usefulness for applications. It is worth noting
that in the case of ErPdBi the relatively low thermal conductivity results
in figure of merit of approximately 0.08 at 500~K\cite{sekimoto}. This value
of $ZT$ is similar to those found for doped 3$d$-electron transition
metal-based half-Heusler phases (see Refs.\onlinecite{hohl,shen}).

\begin{figure}[b]
\centering
\includegraphics[width=0.42\textwidth]{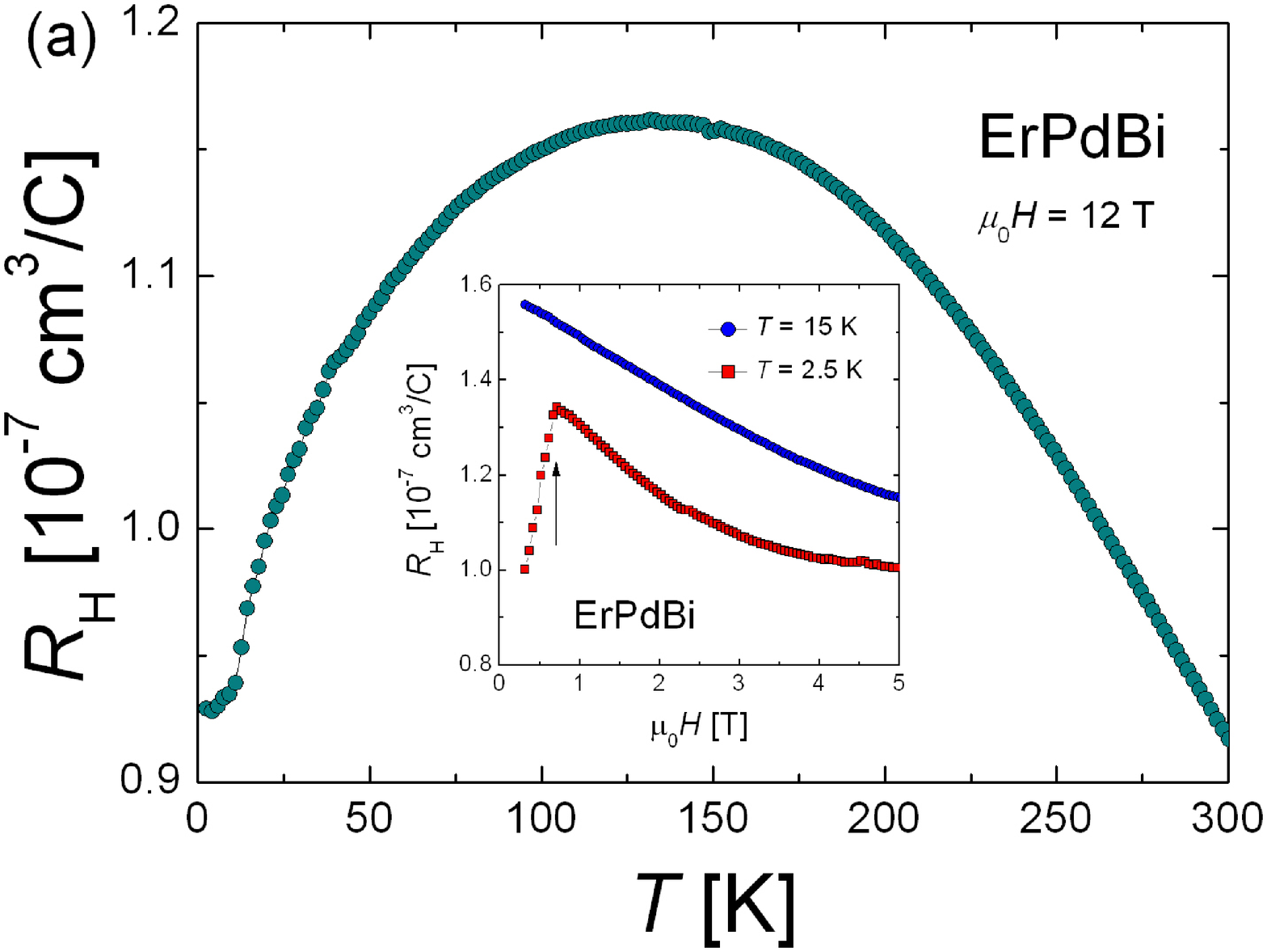}
\includegraphics[width=0.42\textwidth]{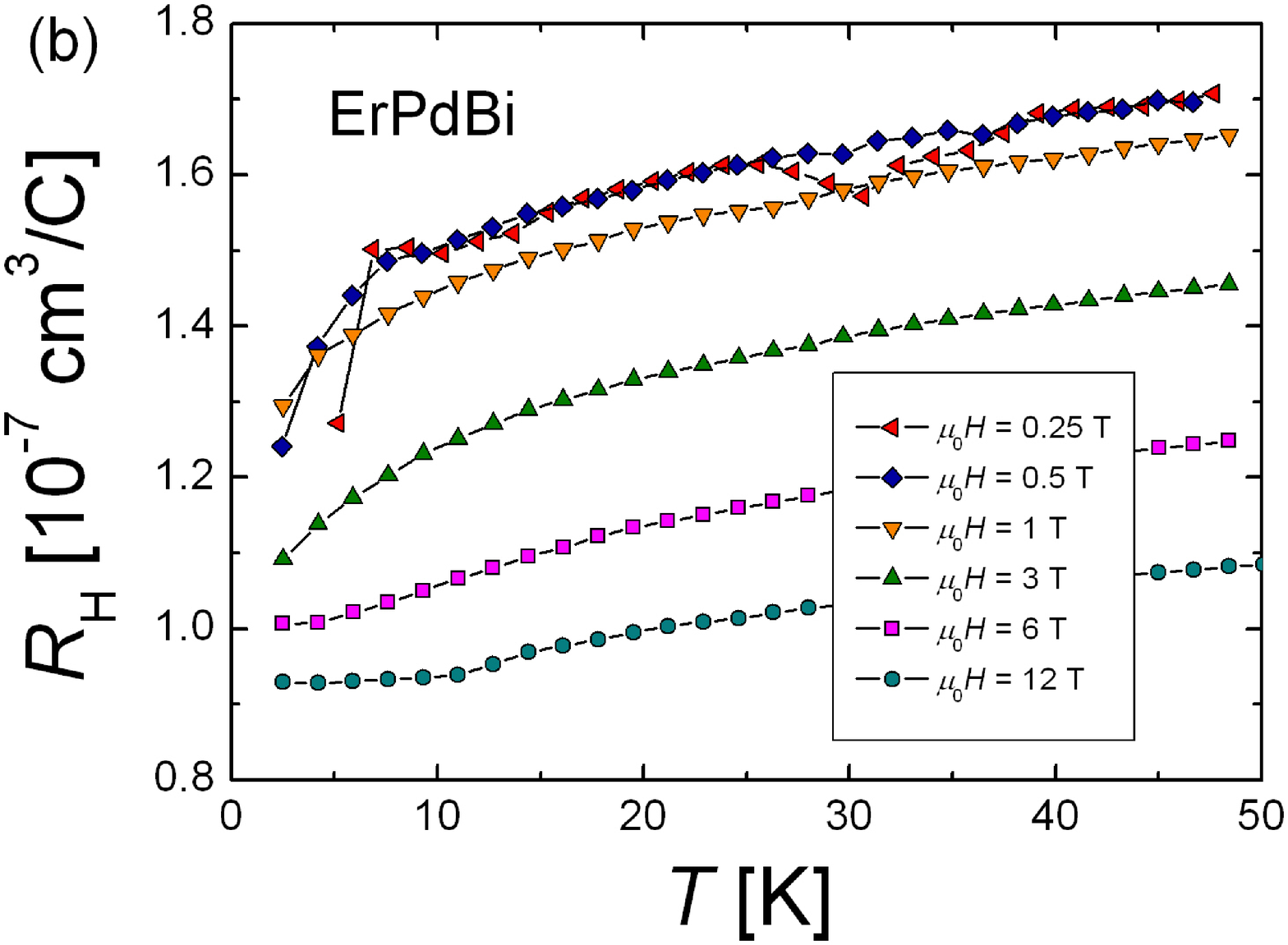}
\caption{(color online) (a) Temperature dependence of the Hall coefficient of ErPdBi. Inset: $R_{H}(B)$ curves obtained at 2.5 and 15~K. (b) Low temperature dependence of $R_{H}$ measured in several different magnetic fields.}\label{zt}
\end{figure}

The temperature dependence of the Hall coefficient of ErPdBi is shown in Fig.~5a. The positive sign of $R_{H}$ throughout the entire temperature range indicates that holes dominate the electrical and heat transport. This is fully consistent with the positive sign of the Seebeck coefficient observed in this compound. As shown in Fig.~5b and in the inset to Fig.~5a, $R_{H}$ exhibits strong temperature and magnetic field dependencies. This finding further supports the conjecture on the complex electronic structure in ErPdBi, where the electron and hole bands have temperature and magnetic field dependent carrier concentrations and mobilities. At the lowest temperature, the Hall coefficient is about 0.93~$\times$~10$^{-7}$~$m^{3}/C$. It is almost three times smaller than that found in the narrow band semiconductor ErPdSb (Ref. \onlinecite{KGprb2}) but still 10$^{2}$-10$^{4}$ times larger than $R_{H}$ observed in simple metals. Within the single band model one may estimate the carrier concentration $n_{H}$ and the Hall mobility $\mu_{H}$ to be of the order of 10$^{19}$~cm$^{-3}$ and 100~cm$^{2}$V$^{-1}$s$^{-1}$, respectively. These values were found to be strongly temperature dependent (not shown). The value of $n_{H}$, due to the crude approximation that neglects semimetallic character of the compound, can be considered as an upper limit of the real concentration in this material. It is about an order of magnitude smaller than the one derived from the thermoelectric data, yet close to $n_{H}$ reported for semimetallic compounds \textit{M}NiSn (\textit{M}~=~Hf, Zr, Ti)\cite{aliev3,aliev4}. In turn, the estimated mobility is much lower ($\sim$~8 times) than the bulk mobility obtained for the topological insulator Bi$_{2}$Te$_{3}$\cite{qu}. It is also worth to mention that for the latter compound the surface mobility reaches as large magnitude as 10000~cm$^{2}$V$^{-1}$s$^{-1}$ (see Ref. \onlinecite{qu}).

\section{Summary and conclusions}

The bismuthides \textit{R}PdBi (\textit{R}~=~Er, Ho, Gd, Y, Dy, Nd) crystalize in the cubic MgAgAs-type structure. These compounds, except diamagnetic
YPdBi, exhibit localized magnetism due to $R^{3+}$ ions. Most of them order antiferromagnetically at low temperatures. All the samples studied showed the electrical conductivity reminiscent of semimetals or narrow gap semiconductors. Their thermoelectric power behaves in a manner typical for semimetals with holes as majority carriers. Moreover, for ErPdBi, the Hall effect study revealed a complex electronic structure with multiple electron and hole bands and different temperature and magnetic field variations of the carrier concentrations and their mobilities.

The results obtained for the \textit{R}PdBi compounds seem to be fully compatible with the theoretical predictions for topological insulators.\cite{NM1,NM2} In particular, a non-trivial zero-gap semiconducting state has been postulated for YPdBi, with relatively small topological band inversion strength, whereas a narrow-gap semiconducting state has been predicted for YPdSb (Ref. \onlinecite{NM2}). These theoretical results agree well with the experimental data, presented in this work and in Ref. \onlinecite{KGprb2}. It is worth noting that YPdBi (maybe also some other \textit{R}PdBi compounds) is located very close to the inversion transition\cite{NM2}, and hence it should be relatively easy to lift the degeneracy and open an inverted gap up at the Fermi level by small distortion of the cubic structure via alloying or external pressure. This intriguing hypothesis should motivate further studies in the field. In the latter context, one should stress that even though the \textit{R}PdBi phases exhibit significant bulk carrier densities, the existence of topological state in these compounds cannot be ruled out. Hitherto studied materials such as Bi$_{2}$Se$_{3}$ or Bi$_{1-x}$Sb$_{x}$ have been predicted to be topological insulators only if they are perfectly crystalline. Real samples always have impurities and defects causing them to be not truly insulating, but to possess a finite bulk carrier density. Even materials which display a bulk insulating state, in surface sensitive experiments such as STM or ARPES, still show finite bulk carrier density in transport measurements\cite{a1,a2}. Clearly, more investigations in this research field are required on well-defined samples with the highest quality, involving wide range of different experimental techniques.

Last but not least, all the \textit{R}PdBi compounds studied, and especially DyPdBi, exhibit the large power factor coefficients $PF$~=~6-20~$\mu$Wcm$^{-1}$K$^{-2}$. This result should motivate further work towards determining their figure of merit coefficient $ZT$, which measures the thermoelectric efficiency in real applications.

\begin{acknowledgments}
We are indebted to U. Burkhardt and H. Borrmann for metallographic and x-ray powder analysis of the samples. We thank T.~Durakiewicz and J.~Lashley for valuable discussion.

\end{acknowledgments}

\end{document}